# Three-Dimensional Anisotropic Magnetoresistance in the Dirac Node-Line Material ZrSiSe

Haiyang Pan[1,a)], Bingbing Tong[2,3,a)], Jihai Yu[1], Jue Wang[1], Dongzhi Fu[1], Shuai Zhang[1], Bin Wu[1], Xiangang Wan[1], Chi Zhang[2,3,], Xuefeng Wang[4, b)], Fengqi Song[1,b)]

[1]National Laboratory of Solid State Microstructures, Collaborative Innovation Center of Advanced Microstructures, School of Physics, Nanjing University, Nanjing 210093, P. R. China

[2]International Center for Quantum Materials, Peking University, Beijing, 100871, China

[3]Collaborative Innovation Center of Quantum Matter, Beijing, 100871, China

[4]National Laboratory of Solid State Microstructures, Collaborative Innovation Center of Advanced Microstructures, School of Electronic Science and Engineering, Nanjing University, Nanjing 210093, P. R. China

---

[a)]H. Pan and B. Tong contributed equally to this work.

[b)]Authors to whom correspondence should be addressed. Electronic mails: xfwang@nju.edu.cn (X.W.) and songfengqi@nju.edu.cn (F.S.)

.

**Abstract**

The family of materials defined as ZrSiX (X = S, Se, Te) has been established as Dirac node-line semimetals, and subsequent study is urgent to exploit the promising application of unusual magnetoresistance property. In this work, we systematically investigated the anisotropic magnetoresistance in the newly-discovered Dirac node-line material ZrSiSe. By applying a magnetic field of 3 T by a vector field, the three-dimensional (3D) magnetoresistance (MR) shows strong anisotropy. The MR ratio of maximum and minimum directions can reach 7 at 3 T and keeps increasing at the higher magnetic field. The anisotropic MR forms a butterfly-shaped curve, which indicates the quasi-2D electronic structures. This is further confirmed by the angular-dependent Shubnikov-de Haas (SdH) oscillations. The first-principles calculations establish the quasi-2D tubular-shaped Fermi surface near the X point in the Brillouin zone. Our findings shed light on the 3D mapping of MR and the potential applications in magnetic sensors based on ZrSiSe Dirac materials.

Recently, Dirac materials have attracted extensive attentions due to their unusual linear band crossing at fourfold degenerate Dirac points between the valence band and the conduction band in the Brillouin zone (BZ). The extended lines or closed loops of the crossing Dirac points in the BZ will form the interesting Dirac node-line materials (DNLMs).[1-6] The Dirac node-line electronic state contributes to a high carrier density of ~$10^{21}$ cm$^{-3}$,[7,8] significantly larger than those of graphene and other Dirac semimetals.[9-18] PbTaSe$_2$,[19] PtSn$_4$[20] and the *WHM* (*W* = Zr/Hf/La, *H* = Si/Ge/Sn/Sb, *M* = O/S/Se/Te)[21] compounds were theoretically predicted and experimentally proved to be typical DNLMs.[22-28] Particular attention has been paid to materials in the *WHM* family due to the fact that they exhibit the space group of iron-based superconductors and their monolayers are suggested to be two-dimensional (2D) topological insulators.[21] The diamond-shaped Fermi surface hosting line nodes and the symmetry-protected band indicate the formation of the node-line semimetal (NLSM) phase in ZrSiX (X=S, Se and Te) family.[22,23,25,28,29] The nearly electron-hole-compensated carriers result in the high magnetoresistance (MR) ratio[30] as well as the large anisotropic magnetoresistance (AMR).[31-33] The AMR is important to realize magnetic sensors in spintronic applications, which offers the possibility to explore the technological applications of NLSMs.[34]

Up to now, the AMR has been observed in many topological materials, such as WTe$_2$,[35,36] NbP[37] and LaBi.[38] The magnetoresistance (MR) of fixed magnetic field value in ZrSiX (X=S, Se, Te) shows the clear butterfly-shaped anisotropy when changing the direction of magnetic field.[7,30,31,33] However, all these AMR measurements were carried out in a 2D plane, limiting insight into the complete AMR effect in these discovered Dirac materials. It remains yet unexplored that 3D mapping of MR in NLSMs, which is of vital importance for magnetic sensor device applications.

In this article, we systematically investigated the AMR in the newly-discovered Dirac NLSM ZrSiSe. The 3D mapping of the MR exhibits the strong anisotropy. The minimum and maximum of MR is observed when the magnetic field sweeps along [100] and [011] family directions, respectively. The AMR curves in *bc*- and *ac*-planes show the butterfly shape, indicating the quasi-2D electronic structure. The 2-fold and 4-fold symmetries of the butterfly-shaped MR are robust at the high magnetic field. The angular-dependent Shubnikov-de Haas (SdH) oscillations further depict a quasi-2D Fermi pocket with the frequency of 210 T. This establishes the tubular-shaped Fermi surface near the X point in the Brillouin zone. The AMR behavior is associated with the quasi-2D Fermi surface structure. Our work provides a route to realize magnetic sensor devices based on the anisotropic Fermi surface of topological Dirac NLSMs.

Figure 1(a) shows the X-ray diffraction image of the exposed plate-like surface of a ZrSiSe crystal. The presence of sharp (00L) peaks can be attributed to the (001) plane (*c*-plane) of the plate-like surface and the high crystallinity of ZrSiSe crystals. As shown in Figure 1(b), the temperature-dependent resistivity at zero magnetic field exhibits metallic behavior and almost reaches saturation below 50 K. When a magnetic field is applied along *c*-axis, the temperature-dependent resistivity increases at low temperature and undergoes a drastic enhancement ($B >3$ T), which results a minimum resistivity. This behavior is very similar with the metal to insulator transition, and is observed in many topological semimetal materials with ultrahigh mobility and large MR, such as NbP,[37] WTe$_2$[36,39,40] and LaBi.[38] The resistivity saturate plateau also appears at low temperature with different external magnetic fields. To analysis the resistivity transition, we plot the $T_m$, where the resistivity reduces to minimum, and $T_i$, where the resistivity reaches to a saturate plateau, versus magnetic field in Figure 1(c). It's evident that $T_m$ keeps monotonous increase as magnetic field is increased. Otherwise, $T_i$ is almost unchanged from low to high magnetic field. This temperature-dependent magnetic field relationship is a common intrinsic property of topological semimetals.[41]

As previous reported, the MR of fixed magnetic field values in ZrSiX (X=S, Se, Te) shows clear butterfly shape anisotropy when changing the direction of magnetic field.[7,30,31,42] Due to the limit configuration of the solenoid magnets, all these reported MR measurements of different magnetic field directions were performed in a 2D plane and the study of MR in other directions is still absent. To further reveal the anisotropic property of MR, we carried out the 3D space MR measurement of 3 T. Figure 2(a) is the schematic of 3D space MR measurement. The current *I* is always applied along *a*-axis during the MR measurements. The magnetic field direction of 3D space is described by *α* (the angle between the magnetic field and *c*-plane) and *β* (the angle between the current and the *c*-plane projection direction of *B*). The MR=[*ρ(B)*-*ρ(0)*]/*ρ(0)* is commonly used to illustrate the change amplitude of resistivity due to the applied magnetic field *B*. Figure 2(b) shows the 3D space MR mapping of 3 T at 1.7 K in ZrSiSe bulk crystal. All the MR is performed symmetry treatment by averaging over the two opposite magnetic field directions to rule out the Hall signal. From the MR mapping distribution, the anisotropic feature of MR can be clearly seen. Corresponding color plot of the 3D space MR is displayed in Figure 2(c). At a certain *β*, the angular-dependent MR almost exhibits butterfly shape (Figure 2(d)) with continuous changing *α*, and the maximum values of MR appear near *α* =45º,135º, 225º and 315º directions. As *β* increases, MR at *α* =45º family directions keep to increase and reach the maximum at *β*=90º. To clearly see the angular-dependent MR property in *bc*-plane and *ac*-plane, Figure 2(d) displays the polar plot of angular-dependent MR of 3 T at *β*=0º (black line) and *β*=90º (red line) simultaneously. This reveals that the maximum MR occurs along [011] family directions, and the minimum MR appears along [100] family directions. The ratio of MR at maximum/MR at minimum reaches to 7 at 1.7 K and 3 T. The 3D space MR mapping reveals the strong anisotropy electronic structure in ZrSiSe crystal. The butterfly shape MR of *bc*-plane and *ac*-plane is the same as the MR observed in ZrSiS.[7,30,31,42,43] As reported in ZrSiS, the butterfly shape MR can be regard as 2-fode and 4-fode symmetries,[42] which is very different from the AMR observed in WTe$_2$,[35,36] LaBi[38] and Bi.[44,45] For the perfect 2D electronic Fermi surface structure, the angular-dependent MR exhibits a typical 2-fold symmetries anisotropy. In contrast, the MR is almost unchanged for the 3D isotropic electronic structure system.[32] Thus, the AMR in ZrSiSe is associated with the quasi-2D nature of the Fermi surface.

To further reveal the physical origin of AMR in ZrSiSe, it's necessary to obtain the Fermi surface structure. The angular-dependent SdH oscillation is an effective method and is widely used to establish the shape of Fermi surface in experiment. We performed the angular-dependent SdH oscillations measurement along two main crystal planes (*bc*-plane and *ac*-plane) to construct the entire Fermi surface in ZrSiSe.

We first measured MR curves at different magnetic field directions along the *bc*-plane as the magnetic field is always perpendicular to the current direction to maintain a constant Lorentz force. The inset of Figure 3(a) shows the MR measurement scheme for the *bc*-plane. The magnetic field direction is indicated by $\varphi$, that is, the angle between the magnetic field and the *c*-axis on the *bc*-plane. The MR measurements with $\varphi$ varying from 0° (*c*-axis) to 90° (*b*-axis) were carried out along the *bc*-plane at 2.3 K. As shown in Figure 3(a), the SdH oscillations can be clearly seen at different magnetic field direction. In order to clearly observe the difference of MR oscillations, we only depict certain MR curves corresponding to different magnetic field directions in Figure 3(a). The other MR curves is displayed in Figure S3 (Supporting information). In the meantime, we also performed the angular-dependent MR measurements under higher values of different magnetic field to study the AMR effect. Figure 3(b) shows the polar plot of angular-dependent MR from 4 T to 14 T at *bc*-plane and 2.3 K. The same as MR curve of 3 T at *bc*-plane, the MR of higher magnetic field also exhibits clear butterfly shape anisotropy. The detail normal plot of the angular-dependent MR is displayed in Figure S4, from which the 2-fode and 4-fode symmetries of butterfly shape can be clearly seen. When the magnetic field increased to 7 T, there are some wiggles appearing at the maximum value of MR. The wiggles become more prominent as the magnetic field value increased to 14 T, which is associated with the strong quantum oscillations of Landau level. As observed in WTe$_2$,[35] the appearance of these additional oscillation wiggles keeps consistent with the observed Shubnikov–de Haas (SdH) oscillations. The symmetries are also kept when the magnetic field is increased to 14 T.

Figure 3(c) displays the amplitude of the angular-dependent SdH oscillations with $\varphi$ varying from 0° (*c*-axis) to 90° (*b*-axis) along the *bc*-plane at 2.3 K. The SdH oscillation amplitude $\Delta\rho$ is obtained from MR measurements (Figure 3(a) and Figure S3) after smooth polynomial background subtraction. For better visibility, the oscillation amplitude is plotted in the form of $\Delta\rho/\rho_0$ (where $\rho_0$ represents the resistivity at zero magnetic field), and the curves are shifted. The oscillation patterns are clearly different for various tilt angles. Figure 3(d) shows the corresponding fast Fourier transformation (FFT) spectra of angular-dependent SdH oscillations. We only observe a frequency oscillation of 210 T when the magnetic field lies along the *c*-axis ($\varphi = 0°$). Otherwise, the oscillation frequency modes are more than one when the magnetic field is applied along the *b*-axis ($\varphi = 90°$) in the *c*-plane. It is evident that the *F*=210 T frequency mode always exists and gradually shifts to 460 T with $\phi$ various from 0º to 90º. In addition, the MR curves of different magnetic field directions with $\phi$ various from 90º to 180º were also measured at *bc*-plane (Figure S5). The MR and FFT spectrum are almost the same as measured results with $\phi$ various from 0º to 90º. The magnetization measurement was also carried out to probe de Haas–van Alphen (dHvA) oscillations. The obtained results in and out of the *c*-plane (Figure S6) are consistent with angular-dependent SdH measurement.

To construct the complete Fermi surface, we also performed angular-dependent MR measurements in the *ac*-plane at 2.7 K. The corresponding measurement scheme is shown in the inset of Figure 4(a). The magnetic field is parallel to the current direction when $\theta = 90°$. For clarity, only certain MR curves corresponding to different angles are displayed in Figure 4(a). Due to the effect of the non-uniform Lorentz force, the MR is very small, and SdH oscillations are unobservable as the magnetic field lies close to the *a*-axis. The angular-dependent MR for different magnetic field values are also measured in the *ac*-plane. As shown in Figure 4(b), the polar plot exhibits strong anisotropy, similar to the case of the *bc*-plane [Figure 3(b)]. The Lorentz force has little effect on the electron movement when the magnetic field lies along the *a*-axis, thereby resulting in an almost unchanged MR for different field values. Figure 4(c) shows the oscillation amplitude $\Delta\rho/\rho_0$ with $\theta$ varying from 0° to 80° at 2.7 K. As in the case of SdH oscillations in the *bc*-plane, the different oscillation patterns are evident at various tilt angles. The corresponding FFT spectra are displayed in Figure 4(d), and the shift of the $F = 210$ T frequency oscillation can be clearly seen with change in $\theta$. The MR curves and corresponding FFT spectrum with $\theta$ various from 90º to 180º were also measured at *ac*-plane (Figure S8). The coexistence of 2-fode and 4-fode symmetries of butterfly shape MR in *ac*-plane (Figure S9) can be clearly seen at the low and high magnetic field of angular-dependent MR. The higher magnetic field almost has no effect on the 2-fode and 4-fode symmetries of butterfly shape AMR at *bc*-plane and *ac*-plane, which suggests that ZrSiSe material is suited to sense the direction of high magnetic field.

Combined with the angular-dependent MR in *bc*-plane and *ac*-plane, we can estimate the MR ratio of maximum ([011] family directions) and minimum ([100] family directions). As shown in Figure 5(a), the MR ratio keeps linear increase when the magnetic field elevates to high value. There is even no saturate signal as the magnetic field increase to 14 T. This relationship can be used to deduce the magnetic field strength in future magnetic sensor device application. The angular-dependent frequency of 210 T SdH oscillations in *bc*-plane and *ac*-plane are plotted in Figure 5(b) and Figure 5(c), respectively. It is evident that the $F = 210$ T frequency mode always exists and gradually increase as $\varphi$ and $\theta$ varies from 0° to 90°. This result indicates that the cross-sectional area of the Fermi surface corresponding to $F = 210$ T does not present the complete 3D case. For the 3D Fermi surface, the corresponding frequency is almost unchanged for different tilt angles of the magnetic field.[7,32,33] For the 2D Fermi surface, the frequency and its corresponding cross-sectional areas shift following the $1/\cos(\beta)$ law at different tilt angles of magnetic field, where $\beta$ denotes the tilt angle between the magnetic field and the normal direction of 2D surface.[7,32] Based on these analyses, we can conclude that 3D and 2D characteristics coexist in the Fermi surface of ZrSiSe. The formula[33] $F = F_1/\cos(t \times \beta)$ can be used to fit the $F = 210$ T frequency mode, where $F_1$ and $t$ denote fitting parameters. Parameter $t$ is determined by the dimensionality of the Fermi surface, and $t = 0$ and 1 for the 3D and 2D cases, respectively. The tilt angle $\beta$ is equal to $\varphi$ or $\theta$, namely, the angle between magnetic field and the *c*-axis of the ZrSiSe crystal in *bc*-plane or *ac*-plane. The blue solid line in Figure 5(b) and Figure 5(c) show the perfect fitting result, from which the fitting parameters in *bc*-plane ($F_1 = 210$ T and $t = 0.70$) and *ac*-plane ($F_1 = 210$ T, $t = 0.70$) are obtained. The value of the dimensionality parameter ($t = 0.70$ and 0.74) locates between 0 and 1, which confirms the coexistence of 2D and 3D characteristics for the $F = 210$ T frequency mode. In addition, a low oscillation frequency near $F = 30$ T is also observed with almost no shift in *bc*-plane and *ac*-plane, which indicates the 3D property of the Fermi surface. Other than the high (210 T) and low (30 T) oscillation frequency modes, there also exist many other frequency modes at large scale angles when the magnetic field changes in the *bc*-plane and *ac*-plane. These frequency modes are related to the Fermi pockets near Fermi level.

In order to understand the physical origin of the observed AMR behavior, we performed first-principles density functional theory (DFT) calculations to further investigate the properties of the electronic band structures and the Fermi surface in ZrSiSe. Figure 5(a) shows the calculated band structure of ZrSiSe along various high-symmetry directions in the BZ. The SOC is not considered in our DFT calculations since it has negligible effect on the band structure.[25] Several Dirac cones can be observed near the Fermi level, and Dirac-cone-like features are also visible below the Fermi level, which result is very similar to the calculated band structure of ZrSiS.[22,28,46] These band structure features indicate that electron and hole pockets coexist at the Fermi surface. The observed nonlinear Hall signal (Figure S2) confirms the coexistence of multiple carrier types in the electrical transport. The corresponding Fermi surface is displayed in Figure 6(b). As shown in Figure 6(c), the diamond-shaped Fermi surface formed by the linearly dispersing Dirac cone bands near the Fermi level is clear emerged when observed along the $k_z$ direction. The diamond-shaped Fermi surface is composed of four lens-shaped pockets. In addition, four small quasi-2D tubular-shaped Fermi pockets appear at the corners of the diamond-shaped Fermi surface (X locations in BZ). A similar quasi-2D electrical structure has also been reported in the tubular-shaped Fermi surface of ZrSiS[8,32,33] and the cylindrical Fermi surface of $CeCo_2Ga_8$.[47] The diamond-shaped Fermi surface around Γ and small pockets near X point have been observed via ARPES measurements in ZrSiS and ZrSiSe.[25,28] The cross-sectional area of the observed small Fermi pocket near X point is $\sim 3 \times 10^{-2}$ Å$^{-2}$ in ZrSiSe,[25] consistent with our calculated area order ($10^{-2}$ Å$^{-2}$). With the use of the Onsager relation,[48-50] we can obtain the cross-sectional area of the Fermi surface associated with the $F$ = 210 T frequency oscillation as $S_F$ = 2 × 10$^{-2}$ Å$^{-2}$, which value is very close to the area obtained with ARPES measurements at the X point in the BZ.[25] The cross-sectional area of lens-shaped pockets is evidently larger than that of the tubular-shaped Fermi pockets. This indicates that the $F$ = 210 T frequency oscillation corresponds to the small quasi-2D tubular-shaped pocket at X. Recent high-field SdH-oscillation studies have revealed that the high frequency of 240 T in ZrSiS arises from the petal-like Fermi surface pocket near the X point.[46] In addition, Ali et al. have reported on a high frequency of 243 T in ZrSiS, originating from the tubular-shaped Fermi pocket near the X point.[33] Due to the usual band features of ZrSiSe, the quasi-2D tubular-shaped Fermi pocket near the X point and other multiple band pockets at the Fermi level simultaneously contribute to the electrical transport. This eventually results the AMR behavior in ZrSiSe.

As the Fermi surface is made from Dirac bands, these unusual electrical behaviors, such as unsaturated MR and AMR, are related to the nontrivial Dirac electronic state. Besides, the range of linear Dirac band dispersion (~2 eV) in ZrSiSe is largest in the present discovered Dirac materials. This make ZrSiSe material to be a robust topological system for novel quantum phenomenon study. In addition to the AMR behavior, the low resistivity and high carrier density allow the possible technological applications in the industry.

In conclusion, we first performed the 3D space MR measurement in the Dirac Node-Line material ZrSiSe. The 3D space MR mapping of 3 T and 1.8 K exhibits strong anisotropy. The minimum and maximum of MR is observed when the magnetic field along [100] family directions and [011] family directions respectively. The AMR in the main crystal planes keeps the butterfly shape curves with 2-fode and 4-fode symmetries to 14T. The ratio of MR at maximum/MR at minimum is proportional to the magnetic strength, indicating the stable of AMR effect in ZrSiSe. Through angular-dependent SdH oscillations and band calculations, the strong AMR behavior in ZrSiSe is associated with its anisotropy Fermi surface. The 3D anisotropy property of ZrSiSe provides the opportunity to realize the magnetic sensor devices and offer the possibility to explore the technological applications of topological semimetal materials.

**Methods**

**Crystal growth and characterization:** Single crystals of ZrSiSe were grown using the chemical vapor transport method with iodine (I$_2$) as the transport agent. Stoichiometric mixtures of Zr, Si, and Se powder were loaded into a quartz tube with iodine (5 mg/cm$^3$). The length of the quartz tube used for vapor transport was about 18 cm. The quartz tube was evacuated, sealed, and placed in a horizontal tube furnace such that a temperature gradient was created from 950 °C to 850 °C. The end of the quartz tube that contained the reaction powder was located in the high-temperature zone that was heated to 950 °C, and the empty end of the quartz tube was maintained at 850 °C. The quartz tube was kept in the tube furnace for two weeks and then cooled down to room temperature. Finally, plate-like crystals with a metallic luster were obtained at the cold end. The structure of ZrSiSe single crystals was examined by means of a Bruker-D8 ADVANCE X-ray diffractometer (XRD) using Cu-Kα radiation (=1.5406 Å). The XRD data were collected over the 2$\theta$ range of 10°–60°.

**Magnetoresistance measurement:** Four-probe measurements of MR and Hall resistivity were carried out on the Quantum Design Physical Property Measurement System (PPMS-14) system. The 3D space magnetoresistance of 3 T was measured on a vector magnet, on which the magnetic field value of vertical and horizontal directions can be changed simultaneously. The 3D space magnetic field of a fixed value can be realized through rotating the sample around the vertical direction and changing the magnetic field value of vertical and horizontal directions.

**Electronic structure calculations:** Band structure of bulk ZrSiSe were calculated in the framework of density functional theory (DFT) using the WIEN2K code with a full-potential linearized augmented plane-wave. The Perdew-Becke-Ernzerhof (PBE) parameterization of the Generalized Gradient Approximation (GGA) as the exchange-correlation functional is implemented. The irreducible Brillouin zone was sampled by a 15*15*7 mesh of k-points.

**Acknowledgements**

We gratefully acknowledge the financial support of the National Key Projects for Basic Research of China (Grant No: 2013CB922103 and 2014CB921103), the National Program on Key Basic Research Project of China (Grant No: 2017YFA0303200), the National Natural Science Foundation of China (Grant Nos: 91421109, 91622115, 11522432 and 11274003), the Natural Science Foundation of Jiangsu Province (Grant BK20160659, BK20130054), and the Fundamental Research Funds for the Central Universities (Grant Nos: 020414380056, 020414380057).

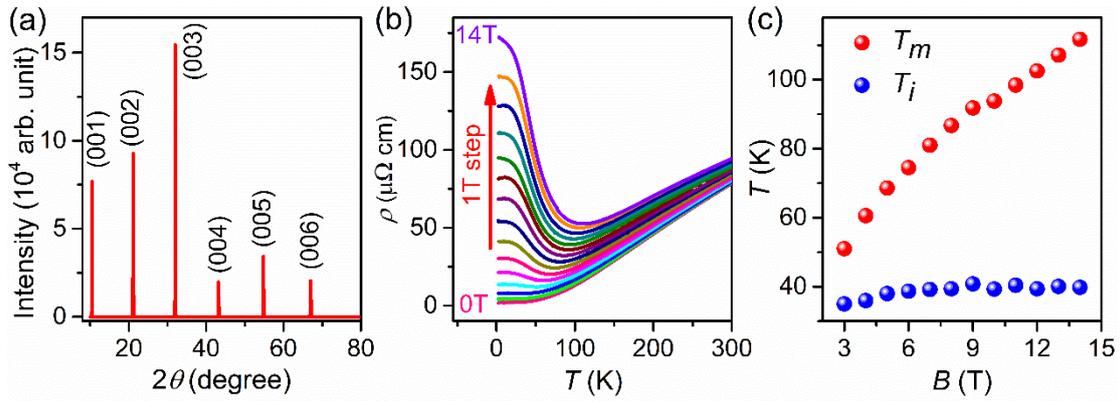

**Figure 1**. X-ray diffraction and temperature-dependent resistivity of ZrSiSe bulk crystals. a) X-ray diffraction patterns of the typical single crystal ZrSiSe. The peak position shows that the sample surface is in (001) plane. b) Temperature-dependent electrical resistivity at 0 T and magnetic field up to 14 T. The magnetic field is applied along [001] direction. c) $T_m$ (red circles) and $T_i$ (blue circles) plotted as a function of magnetic field $B$.

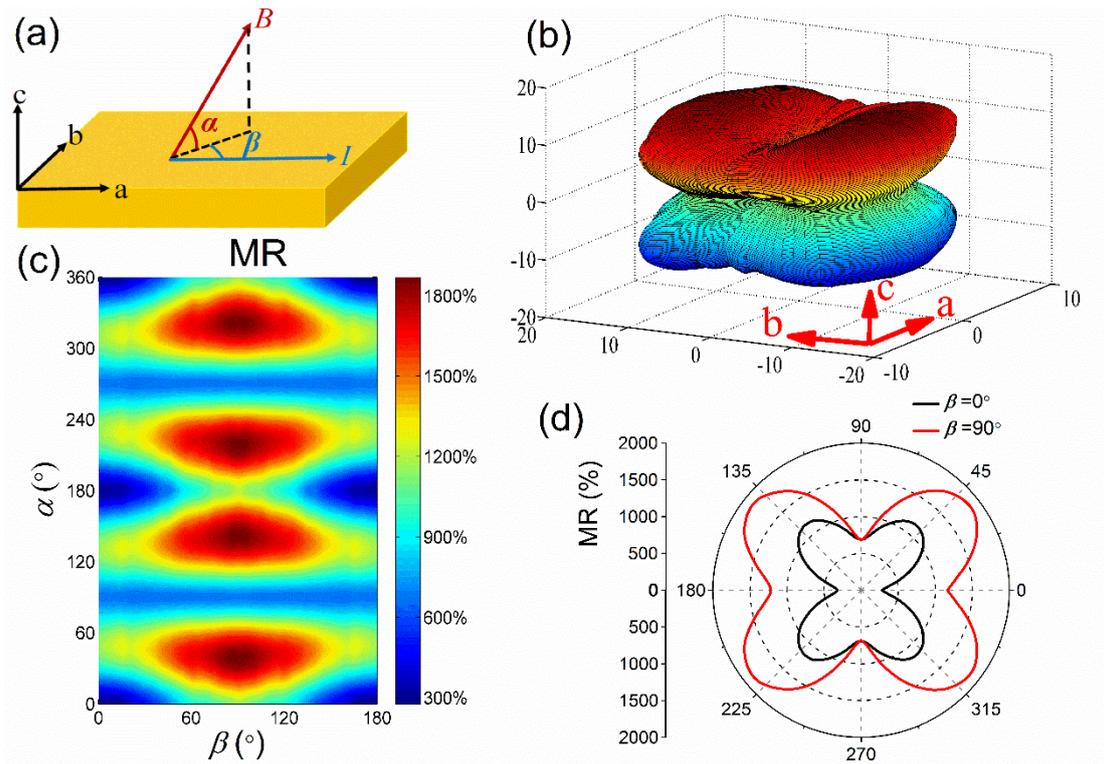

**Figure 2**. 3D space magnetoresistance (MR) measurement of 3 T magnetic field in ZrSiSe crystal at 1.7 K. a) Schematic of 3D space MR measurement. The current is applied along $a$-axis. $\alpha$ is the angle between the magnetic field and $c$- plane, and $\beta$ is the angle between the current and the projection direction of magnetic field $B$ to $c$-plane. b) The 3D space MR mapping at 3 T and 1.7 K. c) Corresponding color plot of the 3D space MR. The right color bar is MR scale. d) The polar plot of angular-dependent MR of 3 T at $\beta=0°$ ($ac$-plane) and $\beta=90°$ ($bc$-plane) respectively.

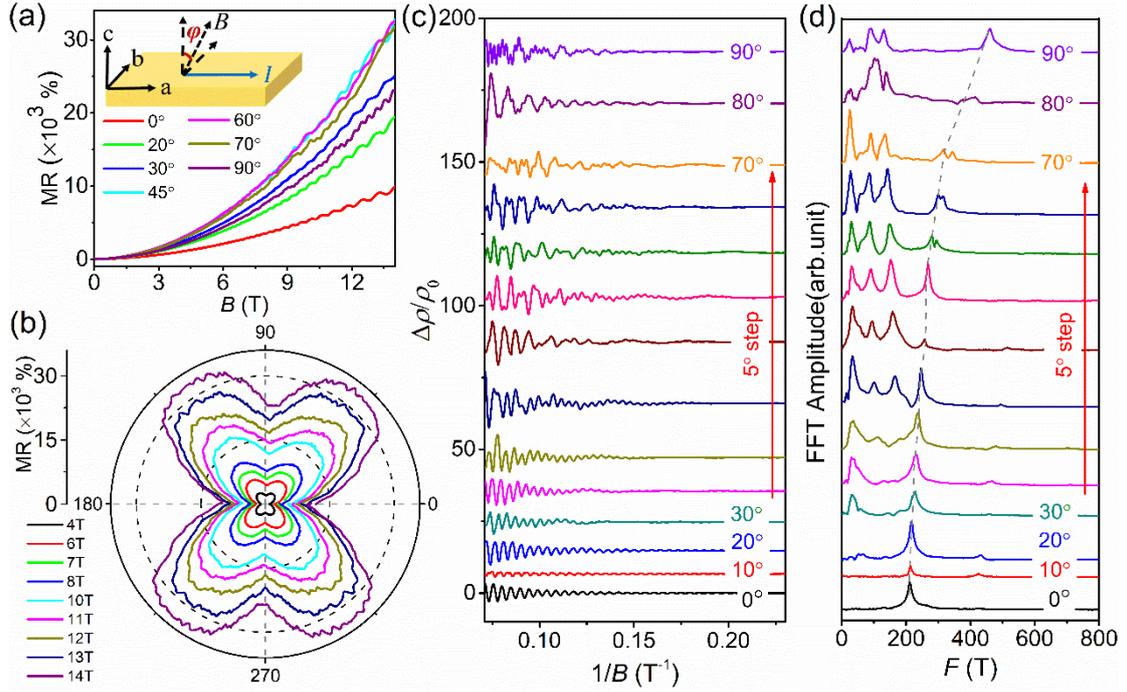

**Figure 3**. Angular-dependent Shubnikov–de Haas (SdH) oscillations of transverse magnetoresistance (MR) at 2.3 K with magnetic field *B* rotated in the *bc*-plane. a) MR measured at different angles with *φ* varying from 0° (*c*-axis) to 90° (*b*-axis). For clarity, only certain MR curves for different angles are displayed. The inset shows the schematic for the field rotation in the *bc*-plane. b) Polar plot of angular-dependent MR for different magnetic fields along the *bc*-plane. c) SdH oscillation amplitude (after polynomial background subtraction) measured with *φ* varying from 0° (*c*-axis) to 90° (*b*-axis). The inset shows the angular-dependent frequency of SdH oscillations. d) Corresponding fast Fourier transform (FFT) amplitude spectra of angular-dependent SdH oscillations in the *bc*-plane.

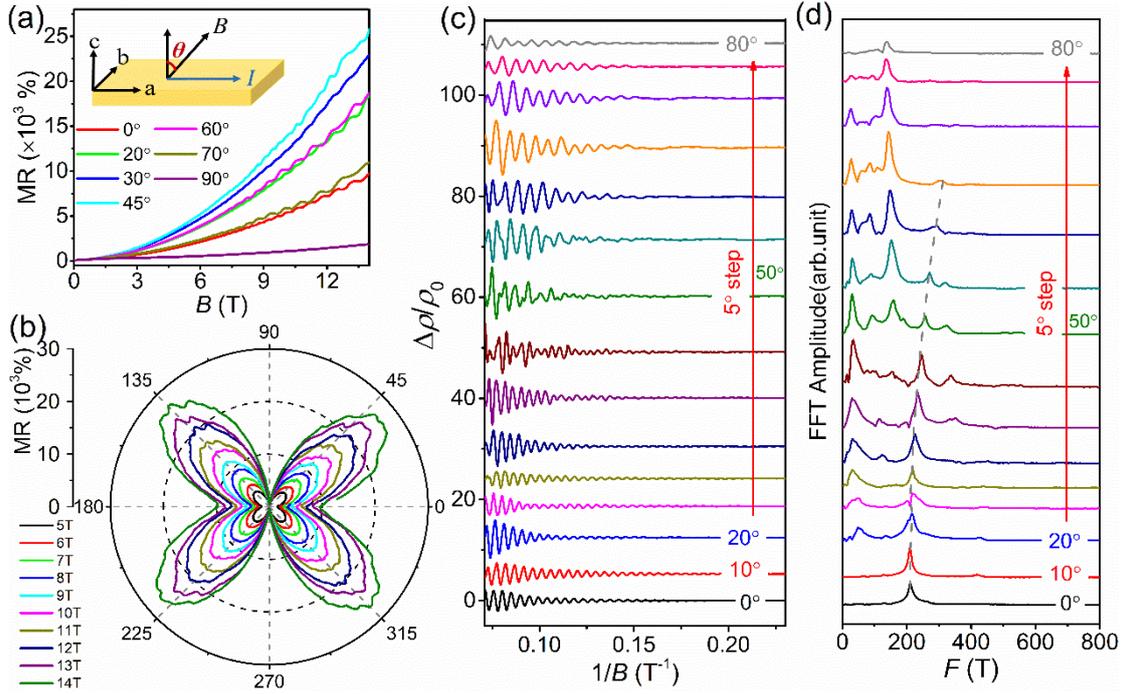

**Figure 4**. Angular-dependent Shubnikov–de Haas (SdH) oscillations of magnetoresistance (MR) at 2.7 K with magnetic field $B$ rotated in the $ac$-plane. a) MR measured at different angles with $\theta$ varying from 0° ($c$-axis) to 80° (near $a$-axis). For clarity, only certain magnetoresistance curves for different angles are displayed. The inset shows the schematic for the field rotation in the $ac$-plane. b) Polar plot of angular-dependent MR for different magnetic fields along the $ac$-plane. c) SdH oscillation amplitude (after polynomial background subtraction) measured with $\theta$ varying from 0° to 80°. The inset shows the angular-dependent frequency of SdH oscillations. d) Corresponding fast Fourier transform (FFT) amplitude spectra of angular-dependent SdH oscillations in the $ac$-plane.

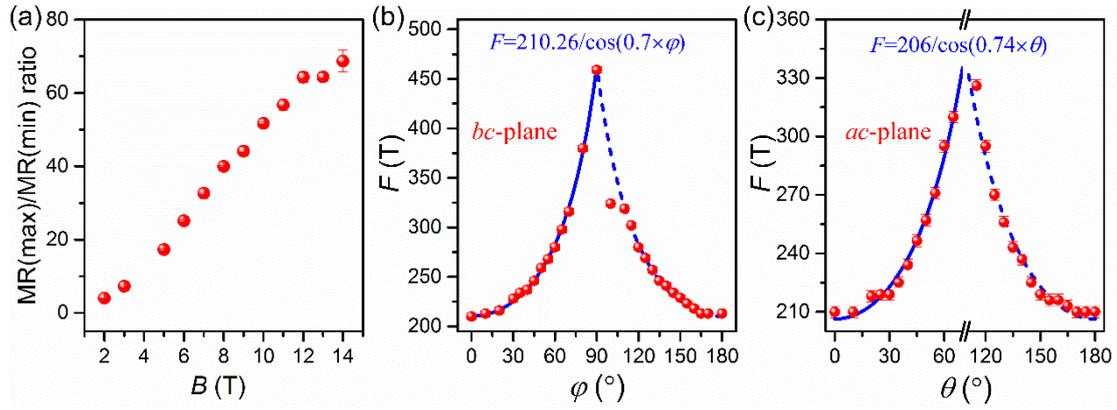

**Figure 5**. The ratio of MR at maximum/MR at minimum and fermiology of ZrSiSe. a) The ratio of MR at maximum/MR at minimum versus magnetic field *B*. b) Angular-dependent SdH oscillations frequencies (red circles) in *bc*-plane. c) Angular-dependent SdH oscillations frequencies (red circles) in *ac*-plane. The blue solid line and the blue dashed lines are the frequency fitted from $F = F_1/\cos(t \times \beta)$ and $F = F_1/\cos[t \times (180°-\beta)]$ respectively. $\beta = \varphi$ (*bc*-plane) or $\theta$ (*ac*-plane).

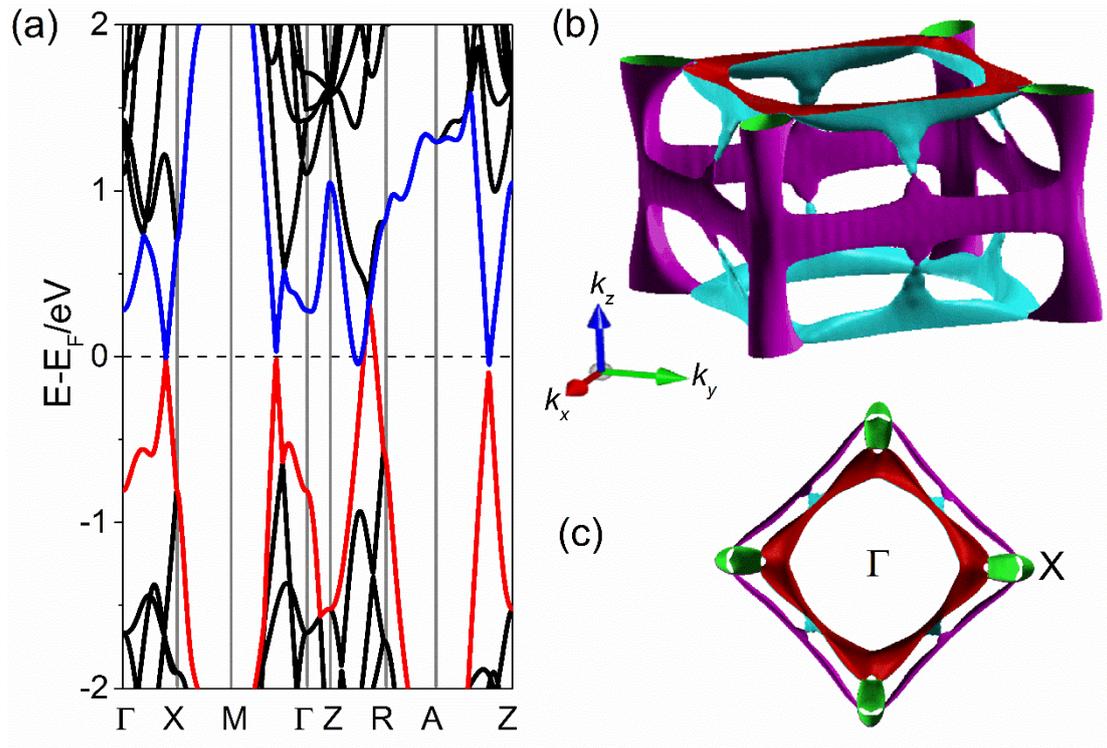

**Figure 6**. Calculated band structure and Fermi surface of bulk ZrSiSe. a) Calculated band structure along various high-symmetry directions without considering spin–orbit coupling. b) Corresponding 3D Fermi surfaces of ZrSiSe in reciprocal space. c) Detailed Fermi surface as observed along the $k_z$ direction.